\documentclass[12pt]{iopart}

\usepackage{graphicx}
\usepackage{natbib}
\usepackage{rotating}

\newcommand{\be}{\begin{equation}}
\newcommand{\ee}{\end{equation}}

\newcommand{\ba}{\begin{eqnarray}}
\newcommand{\ea}{\end{eqnarray}}
\newcommand{\om}{\omega}

\renewcommand\etal{\textit{et al.}}
\newcommand\eg{\textit{e.g.\ }}
\newcommand\cf{\textit{cf.\ }}

\newcommand\apj{{ApJ}}
\newcommand\apjl{{ApJ}}

\newcommand\apss{{Ap\&SS}}
\newcommand\aap{{A\&A}}

\newcommand\mnras{{MNRAS}}

\newcommand\prd{{Phys.~Rev.~D}}

\newcommand\nat{{Nature}}

\begin{document}
\date{\today}  
\title{THE ELECTROMAGNETIC MODEL OF GAMMA RAY BURSTS}
\author{M. LYUTIKOV
}
\address{University of British Columbia, 6224 Agricultural Road,
Vancouver, BC, V6T 1Z1, Canada\\
  Department of Physics and  Astronomy, University of Rochester,
   Bausch and  Lomb Hall,
    P.O. Box 270171,
     600 Wilson Boulevard,
      Rochester, NY 14627-0171, USA }
\begin{abstract}
I describe electromagnetic model  of gamma ray bursts and  contrast its
main properties  and predictions 
with hydrodynamic fireball  model and its 
magnetohydrodynamical extension. 
 The  electromagnetic model assumes that  rotational energy of a relativistic,  stellar-mass
central source (black-hole--accretion disk system or fast rotating 
neutron star)
 is converted into magnetic energy  through unipolar dynamo mechanism,
propagated to large
distances in a form of relativistic,  subsonic,
Poynting flux-dominated wind
and is  dissipated directly into emitting particles through current-driven
instabilities.
Thus, there is no conversion back and forth
between internal and bulk energies as in the case of  fireball model.
Collimating effects of magnetic hoop stresses lead to strongly non-spherical
expansion and formation of jets.
Long and short GRBs may develop in a qualitatively similar way, except that
in case of long burst
ejecta expansion has a relatively short, non-relativistic, strongly dissipative
stage inside the star.
Electromagnetic and fireball models (as well as strongly and weakly magnetized
 fireballs) lead to different  early  afterglow dynamics, 
  before deceleration time. Finally,
  I discuss the models in view of  latest 
  observational data in the 
   Swift era.
\end{abstract}

\maketitle

\section{Short introduction}

Gamma Ray Bursts (GRBs) are conventionally divided into two classes, short-hard and
 long-soft, distinguished by their duration (with a division near $\sim 2$ sec)
 and spectrum hardness \citep{kmf+93}.
Detection 
of Type Ic supernovae nearly coincidence with long GRBs unambiguously linked them
 with deaths of massive stars \citep{smg+03,Hjorth}.
 Studies of the host galaxies 
 of long GRBs, which turned out to be actively star-forming,
 further strengthens this association \citep{Djorgovski}.
 Recent progress
in observations of short bursts showed that
on one hand they show qualitatively similar 
afterglow behavior  (but without any supernovae signature) while on the other
hand their energetics was two  to four orders of magnitude smaller and they
are preferentially (at the moment of writing three out of four) 
associated with older stellar population
\citep{Gehr05,Prochaska05,Villasenor,Covi05a,Rett05,fox}.
These indirect evidences are 
consistent with formation of short GRBs in compact star mergers (double 
neutron stars or black holes--neutron star binaries)
and formation of a black hole \citep[\eg][]{ross03,aloy05}.

\section{Short and long GRBs}

Association of two types of GRBs with different astronomical objects is somewhat surprising given their apparent similarity (perhaps less surprising
in view of the fact that some short GRBs may be associated with nearby SGRs).
One possible reason is that though short and long GRBs occur in different
astrophysical setting, their appearance is governed by similar 
physical process related to formation and early evolution of stellar mass
relativistic compact object. 
[Similarities of  temporal and spectral
properties of the first 2 seconds
in long bursts and short bursts \citep{Nakar02,Ghirlanda03} may be an indication 
of this.]
But then one expects that 
during merger of, \eg, two neutron stars the resulting black hole has large
angular momentum and thus can potentially release much more energy than observed
(one  can invoke different efficiencies, but a naive guess would be
that it's harder to extract and propagate energy from a compact object 
inside a stellar core, contrary to observations).

So 
why short GRBs are so under-energetic if compared with long ones?
One  possibility is that 
presence of a disk
is a necessary condition for extracting energy from a black hole, so after disk disappears
energy extraction stops. In this case
the energy and  angular 
momentum that will power a GRB outflow effectively come not from the
central black hole but from 
the surrounding disk (in a sense that its the energy and/or lifetime  of  the disk 
and not the energy in the black hole
that determines the resulting energy of a GRB outflow)
\citep[for related  discussion  see also][]{vanputten}.

In case of neutron star mergers,
black hole forms fairly early, while the mass of the accretion disk is small,
$\leq 0.1 M_\odot$, with short viscous time scales, $\sim 0.1-1$ sec
\cite[\eg][]{Ruffert}. On the other hand, 
a black hole inside a collapsing core of a massive star may 
 accrete several solar masses of material \citep[\eg][]{mcfad01}
 (at any given moment the mass of the disk is small, but large amount
 of mass, $\sim 1-10 M_\odot$ passes through the disk during accretion).
In addition, amount of the rotational energy stored in case of core collapse
depends on core rotation before explosion (which, in turn, depends  on metallicity
through wind angular momentum loss \cite{woosleyheger}), 
resulting in broad spread of rotational energies.

\section{Principal issues: electromagnetic and fireball  models}

In this contribution we describe 
electromagnetic model of GRBs which assumes that the energy that will power
a GRB comes from rotational kinetic energy of a central source.
The energy is extracted through magnetic field, which can  be generated
by a local dynamo mechanisms
\citep[\eg][]{Hawley,Hawley05,prog03,Gammie}.
As  argued above, the GRB energy should then be related  not to the total
rotational energy of a  central black hole
 but to  the disk around it. Another possibility for long bursts 
is formation of a ''millisecond magnetar'', a
fast rotating strongly magnetized protoneutron star.

Whether magnetic fields  play an important dynamical role
at any stage in the outflow remains, in our view, one of the principal issues  
in  GRBs physics.
Currently, the overwhelming point of view,
 advocated by the fireball model (FBM)
  is that
  magnetic fields do not play any major dynamical role (except, perhaps, at a very early stage;
after which  fields are dissipated quickly).
  FBM advocates that in the emission region magnetic field are
 re-created locally (\eg through development of Weibel instability \citep{med99}),
   with energy density typically much smaller than plasma energy density. Fields are
   small scale, with correlation length $l_c$ much smaller
   than the "horizon" length $l_c \ll R/\Gamma$ ($R$ is the radius of the outflow in the
   laboratory frame and $\Gamma$ is its bulk Lorentz factor).
    Alternative approach, advocated by MHD and electromagnetic models \cite[\eg][]{uso92,bla02,lb03}
  is that there are dynamically important
   large scale fields with "super-horizon" correlation length $l_c \geq R/\Gamma$, which are 
    created at the source and which may play a major role in driving the
    whole outflow in the first place.

To quantify the dynamical
      importance of large scale magnetic fields, it is useful to introduce magnetization parameter
      $\sigma$ as a ratio of Poynting $F_{\rm Poynting} $ to (cold) particle $F_{\rm p}$  fluxes
      (or as a ratio of rest frame energy densities)
      \be
      \sigma= {F_{\rm Poynting} \over F_{\rm p} } =
      {B^2 \over 4 \pi \Gamma \rho c^2} = {b^{\prime 2} \over 8 \pi \rho ' c^2}
      \ee
      where $B$ and $\rho$ are magnetic field and plasma density  in the lab frame, $b^{\prime}$ and $\rho '$
      are magnetic field and plasma density  in the plasma frame (where electric field is zero).
For $\sigma \ll 1$ magnetic fields are dynamically unimportant (this is assumed within 
a framework of a conventional FBM),
while for $\sigma  \geq 1$  magnetic fields start to play important dynamical role. 
For  $\sigma \gg 1$, there is an important qualitative change in the dynamical
behavior of the flow at $\sigma_{crit}= \Gamma^2/2$.
 For   $\sigma< \sigma_{crit}$ the flow is
super-Alfvenic, while for $\sigma > \sigma_{crit}$ the flow
is sub-Alfvenic (and sub-fastmagnetosonic).
 The different is somewhat analogous to the difference between sub-sonic
and supersonic flows in hydrodynamics. 
  
Thus, depending on the parameter $\sigma $  
 three qualitatively different regimes for expansion of the ejecta may be identified,
which  I will call (i)  fireball model (FBM below) $\sigma \ll 1$, (ii) MHD models 
$ 1 \leq \sigma < \sigma_{crit}$ and  (iii)  electromagnetic 
model (EMM below) $\sigma >  \sigma_{crit}$.
These three possibilities  leads to {\it a qualitatively
different dynamic behavior of flows}.
Let us next describe qualitatively main features
of the models.  (As the FBM and EMM are at the extreme range of $\sigma$ we
discuss those first.)
\footnote{Definitions and discussion below is based  not on the nature of the central object but
 on ejecta content at large distances from 
the central region and  before production of $\gamma$-rays occurs.} 

{\bf Fireball model \citep[FBM, \eg][]{Piran04}}:
 The defining characteristic of the FBM 
 is that at intermediate distances (far from the central source  but before
most energy is transfered to the forward shock) most  energy produced by the central source  is
 carried by the bulk motion of  ions. In  temporal order the transformations of energy are as follows.
 Initially, the  energy that will power the GRB and its afterglow
  is thermalized near the central source,
   so that most of it
   is converted into lepto-photonic plasma. 
    This internal energy is then converted to the bulk motion of  ions,
     and reconverted back into internal at internal shocks; at the same time, 
      small scale magnetic fields are generated.
      The energy of these generated magnetic fields is then used to accelerate leptons via Fermi mechanism
      to highly relativistic
      energies 
      \footnote { The energy that goes 
      to non-thermal particles is {\it electro-magnetic}
      even in the FBM:
      Fermi-type acceleration is done by turbulent EMF associated with fluctuations of magnetic field}. 

      {\bf Electro-magnetic model \citep[EMM,][]{lb03}}.
      The defining characteristic of the electro-magnetic model is that the bulk energy of the flow
      is carried {\it subsonically} 
      by magnetic field. In temporal order  the evolution of the energy proceeds as follows.
      The energy that will power a GRB comes from kinetic rotational energy of the central source
      (millisecond pulsar or BH-disk system). It is then converted to magnetic energy using
      unipolar inductor (like in pulsars), transported to
      large distances in a form of strongly magnetized wind 
and is used to accelerate particle in the emission region. Acceleration of particles
is done via magnetic dissipation (not through shocks).

{\bf MHD model \citep[\eg][]{ds02}}:
In this case most  energy is also curried by magnetic field (similar to  \cite{lb03}),
but the flow is supersonic, similar to FBM. In the current version \citep[][]{ds02},
magnetic field energy is first converted into bulk  motion and then dissipated through
internal shocks,  similar
to FBM. This is, in principle, not necessary, so that 
magnetic field  energy may be dissipated directly into  emitting particles, similar to EMM.

The  principal differences
 between EMM  and  MHD
approaches is that 
MHD-type outflow usually cross fast magnetosonic critical surface
after which moment they become  causally disconnected from their source \citep{gj70}.
Initially the flow  is expanding  freely,  so that
the flow dynamics is determined  by the  internal
structure of the flow. Only  after the flow reaches the terminal velocity, the 
interaction with the mediums becomes important.
Unlike MHD,
 force-free flows  are sub-fastmagnetosonic so no conditions
at the fast critical surface appear. In this case it is the {\it interaction with
boundaries 
 that determines the properties of the flow} (similar to subsonic hydrodynamic
flows).
Thus, the distinctive feature between MHD and force-free flows is whether
the wind becomes fast supersonic (MHD regime, $\sigma < \sigma_{crit}  $) or not
(force-free regime, $\sigma > \sigma_{crit}$). This important difference leads to somewhat 
different dynamics of the flow and can be tested with observations, as discussed
in \S \ref{spectr}.

\section{Source Formation and Energy Release in EMM}

\subsection{Electromagnetic luminosity and currents}

In this section I will describe main ingredients of the EMM, stressing its principal difference 
and  predictions from the FBM. 
EMM assumes that 
the GRB ``prime mover'' is relativistic, fast rotating, near stellar-mass
source. As  discussed above, in
 order to reconcile
energies of short and  long GRBs the  ``prime mover'' should be not the black hole
but the disk around it.
  For numerical estimates I  will assume
 that a central source
generates luminosity $L= L_{50} \times 10^{50} $ erg/s for a time $t_s$, where $t_s\sim 100$ s 
for
long bursts and $t_s\sim 1 $ s for short ones 
(there are indications
that both long and short bursts are powered by the same luminosity, but for different
 time \citep{fox}). The mass of the central source is $\sim 0.01 M_\odot$ for short bursts
and $\sim M_\odot$  for long bursts (we stress again that this is the total  mass passing 
through  the disk, not in 
the black hole). The source is assumed to 
rotate with a  spin frequency $\sim $~kHz.
In addition, it is  assumed that a source possesses a large magnetic field of
    $B_s\sim10^{14}L_{50}^{1/2}$~G. 
If initially the core is fast rotating \cite[\eg][]{woosleyheger},
the total rotational energy  in the disk, $\sim M_{disk} R_{disk}^2 \Omega^2 $,
in case of core collapse
 is much larger,  $\sim 7.9 \times 10^{52} ( M_{disk}/M_\odot)
( R_{disk}/ \times 10^6 {\rm cm})^2 ( \Omega/ 6.28 \times 10^3 {\rm rad/s})^2
$ erg,
than in the case of mergers,  $\sim 7.9 \times 10^{50}$ erg 
with $M_{disk} \sim  0.01 M_\odot$ in the above estimate.
 The source 
is expected to be active for
$t_s \sim E/L \sim 100$ s for longs and $\sim 1 $ s for shorts.

Rotational energy of the central object is extracted by magnetic fields through unipolar
induction mechanism, similar to prevailing model of AGN
jets \citep[\eg][]{Ferrari}. 
Magnetic fields both 
launches the jet  (\eg through Blandford-Znajek mechanism)
 and collimates it by hoop stresses.
  Latest full relativistic MHD
  numerical simulations of accretion disk--black hole systems
   do show formation of the strongly-magnetized axial funnel \cite[\eg][]{Hawley,Gammie}.
Thus, {\it large scale, energetically dominant magnetic fields may be expected in the
launching region of GRB jets.}

Qualitatively,
in the immediate vicinity of the 
 source, the  plasma 
is  separated into two phases: an internal  matter-dominated phase in which
large currents are flowing and  external magnetically dominated phase.
Strong  magnetic fields and  magnetic flux are  
generated in the dense medium (a  disk or  a differentially rotating  neutron star-like
object), while  
 relativistic outflow is generated in the 
magnetically dominated phase. 
In this case   matter loading may be expected to be small
(\eg  analogous to pulsar wind).
 As the source  remains  active for $\sim$ thousand to  million dynamical times the
 flow  will be able
 to settle down quickly to a quasi-steady state evolving slowly as the hole or neutron star
 slows down.
The separation  into matter- and magnetic field-dominated  phases
 is somewhat similar to the Sun, where dynamo operates in tachocline,  deep
below the surface, while magnetic energy, and most importantly magnetic flux,
are  dissipated outside the star.

The key assumption of the model is that
 the dissipation rate at the source  remains low enough so that 
the power continues to be dominated by the electromagnetic component 
(rather than the heat of 
a fireball) well out into the emission 
region. Thus, electrical 
currents flows  all the way out
to the expanding blast wave, rather
than been dissipated  close to the source.
For electromagnetically dominated outflows 
the value of the total current may be related to the total luminosity of the source
\be
\label{eqamp}
I\sim  \sqrt{ L c \over 4 \pi}
\sim3 \times 10^{20}L_{50}^{1/2}\ {\rm A},
\label{VI}
\ee
where  notation $X_n =(X/10^n)$ was adopted.
Under general electromagnetic and relativistic conditions
the  total   impedance  of the source and the emission region is
close to the impedance of free space  ${\cal Z}  \sim  100 \, \Omega$

In case of long burst, associated with collapse of massive stars,
 the source will initially inflate a non-relativistically expanding
 electromagnetic bubble inside the star. 
This magnetized cavity is separated from the outside material by the
(tangential)
contact discontinuity (CD) containing  a surface  Chapman-Ferraro
current. This current  terminates the magnetic field
and completes the circuit that is driven by the source.
On a microphysical level the current is created by the
particle of the surrounding medium completing half a turn in 
the magnetic field of the bubble, so that the 
thickness of the current-currying layer is of the order of  ion gyro-radius.
After the break out the density that controls the ejecta expansion falls down, so that 
expansion becomes relativistic.
In case of short burst, associated with merger of 
two neutron stars, a somewhat  similar process will happen, except  that there is no
non-relativistic stage, so that 
the bubble
is directly inflated in  the circumburst medium.

\subsection{Distribution of current in the wind: structured jet}
\label{Itheta}

\begin{figure}
\includegraphics[width=0.9\linewidth]{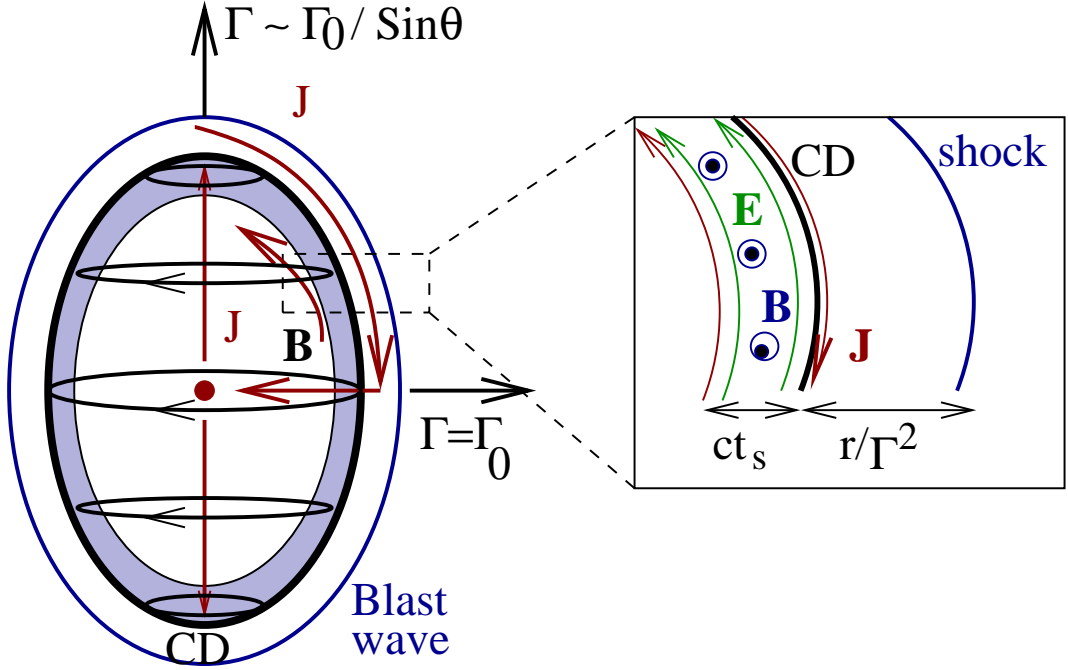}
\caption{Current flow in the electromagnetic bubble.
Current flow mostly along the axis, on the surface of the magnetic shell, along equator
and close-up at the trailing part of the shell. Magnetic shell is preceded by the
forward shock, typically $r/\Gamma^2$ ahead of it.
 Non-sphericity of the shell, which is 
of the order $\sim 1/\Gamma^2$,  is enhanced.}
\label{current}
\end{figure}

The form of expanding bubble depends on lateral distribution of source luminosity, which within 
a framework of EMM mode is given by lateral distribution of current. 
At relativistic stage expansion is nearly ballistic \citep{sha79}
while at a non-relativistic stage  inside a star
 a flow  may be  collimated both through the action of
 magnetic hoop stresses {\it and} interaction with the surrounding gas (see
  \S \ref{formnr}).
One  particular stationary outflow  configuration which captures 
the essential features of the outflow, 
is that the outgoing  current is confined to the poles and the equatorial
plane and closes along the surface of the bubble, Fig. \ref{current}.
This current distribution minimizes the total energy given a total toroidal magnetic flux
and has been advocated in relativistic stationary winds  \citep{HeyN03}.
The magnetic field in the bubble is inversely proportional to  the cylindrical
radius, $B_\phi \propto 1/( r \sin \theta)$. Accompanying this magnetic field is  a poloidal
electrical field so that there is  a near radial
Poynting flux carrying energy away from the source.
Thus, 
as the magnetic field strength is strongest close to the 
symmetry axis, the bubble will expand 
fastest along the polar direction.
The  internal structure of an outflow then corresponds to a ''structured jet''
with $L_\theta \sim \theta^{-2}$ \citep{Lipunov,ros02}, so that
 the central source releases an equal amount of energy per
 decade of $\theta$.

\section{Long GRBs: expansion inside a star }
\label{ssec:nrexpansion}

\subsection{How important is dissipation?}
\label{dissipation}

Some fraction of the central source luminosity
is likely to be dissipated close to the source. The fireball model 
 implicitly assumed that all of  the energy released is quickly
converted into heat, while in the electromagnetic model this does not
happen and the energy flows way from the light cylinder mainly
in the form of an electromagnetic Poynting flux
and the load impedance is located in the emission region.

The issue of dissipation is somewhat complicated as  I discuss next.
On one hand,
somewhat paradoxically,
it becomes harder to convert electromagnetic energy
directly to pair plasma the stronger the magnetic field becomes.
The reasons is that
 the  maximum potential drop that is available for dissipation
 will be  limited by various mechanisms of pair production. Typically,
 after an electron has passed through  a potential difference
  $\Delta V\sim 10^9-10^{12}$ V
  it will  produce an electron-positron
   pair either through the  emission of curvature photon
   or via  inverse Compton scattering. This will be followed by an electromagnetic cascade and
   the   newly born
   pairs will create a charge density that would shut-off the accelerating
   electric field.
 Put another way, the pair density required to supply the electrical 
current and space charge scales linearly with the field strength, while the electromagnetic 
energy density scales as its square. The stronger the field, the more likely it is to persist
into the outflow. 
It is because GRBs are so powerful that the dissipation in the source is probably low.

There is an important caveat to the preceding discussion  which applies  
to very early, nonrelativistic,
stages of bubble expansion in case of long bursts.
Somewhat
similar to  pulsar wind nebulae \citep[\cf][]{rg74},
{\it non-relativistic, ideal, homologous expansion of strongly magnetized nebular which is 
causally disconnected 
from the source and injects
toroidal magnetic field with nearly a speed of light  cannot occur.}
The reason is that magnetic flux and energy are supplied to the 
inflating bubble by a rate that cannot be accommodated in the bubble.
The  rate  of supply is determined by the processes inside the
light cylinder of a newly-formed, compact object, while inflation of the  bubble is controlled by the
external gas density.
(Even if the wind remains subsonic, it is unlikely that processes at the edge of the inflating
bubble would influence  wind generation region near the light cylinder.) 
This leads to the following ''contradiction''. 
Magnetic flux (integrated over the meridional plane) 
is  supplied to the bubble at 
a rate $\dot\Phi\sim 2 Ic$. Similarly, energy 
is  supplied to the bubble at a rate $\dot U_{{\rm EM}}\sim {\cal E}I$.
We can also compute the magnetic flux $\Phi={\cal L} I$ and the energy
stored 
within the bubble $U_{{\rm EM}}={\cal L} I^2/2$, using the self inductance ${\cal L}$.
Let the bubble radius be $R(\theta,t)$ where
$\theta$ 
is the polar angle measured from the symmetry axis defined by the spin 
of the compact object. 
If 
 the magnetic field in the bubble is predominantly toroidal
between cylindrical radii $\varpi_{{\rm min}}$ and $\varpi_{{\rm
max}}=R\sin\theta$, this
is given by
\be
\label{selfinduct}
{\cal L} \sim{\mu_0\over2\pi}\int dz\ln\left({\varpi_{{\max}}\over \varpi_{{\rm
min}}}\right).
\ee
We therefore see that, if the bubble expands 
 sub-relativistically, the rate of supply 
of both flux and energy exceeds the rate at which 
the flux and energy can be stored by a factor 
$\sim[\ln(\varpi_{{\rm max}}/\varpi_{{\rm min}})(dz/dt)/c]^{-1}$
\citep[\cf][]{rg74}. Therefore, for non-relativistic expansion of the bubble
too much flux and  too much energy is generated  by the source.

The way out of the ''paradox'' is that dissipation must become important that 
will destroy some magnetic energy and most importantly eliminate most of the
toroidal flux. 
 Most
of the dissipation
is likely to occur near the axis where the current density is highest and
the susceptibility to current-driven instability is the greatest.
In this case a lateral 
flow of energy will set in carrying the poloidal field lines with it towards the axis.
This, in turn, will leads to the
pile-up of magnetic field near the axis and to faster radial expansion near the axis
(the toothpaste tube effect) \citep{lb03}.

\subsection{Form of the expanding bubble}
\label{formnr}

The dynamics of a non-spherically expanding bubble may be described using the
method of \cite{komp60}.
 Consider  a small section  of non-spherical 
 non-relativistically
expanding 
CD with radius  $R(t, \theta)$.
 The CD   expands under the pressure of magnetic field 
so that  the normal magnetic stress at the bubble surface is balanced
by the ram pressure of the surrounding medium.
At the  spherical polar
angle $\theta$ the CD  propagates  at an angle 
$
\tan \alpha = - { \partial \ln R \over \partial \theta}
$
to the radius vector.
Balancing the pressure inside the bubble $B^2/(8 \pi) = I^2 /( 2 \pi c^2 R^2)$
with the pressure of the shocked plasma gives
\be
\label{bubbleexp}
\left({\partial R\over\partial t}\right)^2= \kappa {I^2(t)\over2\pi R^2\sin^2\theta
\rho(R,\theta)}\left[1+\left({\partial\ln R\over\partial\theta}\right)^2_t\right]
\label{inside}
\ee
where $\kappa $ is a coefficient of the order of unity
which relates the pressure at the CD to the pressure at the forward shock.

Equation (\ref{inside}) shows that non-spherical expansion inside the star is due both
to the anisotropic driving by magnetic fields  {\it and} 
collimating effects of the stellar 
material (the term in parenthesis, which  under certain conditions
tends to 
amplify  non-sphericity).
The rate of expansion of the bubble inside the star
depends upon the density profile  of the stellar envelope and the time evolution of the
luminosity (or, equivalently, of the
current $I(t)$). 
For a given dependence $\rho(R, \theta)$ and $I(t)$ Eq. (\ref{inside})  determines the 
velocity of the CD. Generally solutions will be strongly elongated along the
axis. A simple analytical solution 
 for $I,\, \rho \sim const$ is  
\be
 R (t,\theta) = \left( { 2 \over \pi} {I^2 \over \rho c^2 } \right)^{1/4}{  \sqrt{t} \over 
\sin \theta}.
\ee
(current is related to the luminosity by Eq. (\ref{VI}).
Qualitatively, the bubble and the forward shock will cross
the iron core ($r_c\sim 2.5 \times 10^8$ cm) in $t \sim r_c \sqrt{\rho} / B(r_c)
\sim .3 $ sec, short enough to produce an ample supply of $^{56}Ni$ \citep{Woosl02}.
 If  $M(R)$  is the stellar mass external to radius $R$, then 
 the  breakout  time  is
\be
\label{eqexp}
t_{{\rm breakout }}(\theta) \sim { 1}  \theta_{-1}^2
\left({M\over M_\odot}\right)^{1/2}\left({R\over R_\odot}\right)^{1/2}
L_{50}^{-1/2}~{\rm s}
\ee
The electromagnetic bubble can be confined equatorially
by the star for the duration of the burst  $t_{{\rm breakout }}(\pi/2) \sim 100 $s 
and will expand non-relativistically as
we have assumed. However the expansion along the axis proceeds
on a short time scale and  breakout should occur
early in the burst. 

Thus, along the jet axis
non-relativistic expansion  lasts  for several seconds, which is  much
shorter than the burst duration. 
After breakout the flow
quickly becomes relativistic so there is no need for dissipation anymore.
Thus, relative fraction of dissipated energy is small,
so that overall the flow magnetization remains large.
Most of the dissipation described above will result in creation
of lepto-photonic plasma, which decouples after photosphere, so that the remaining
flow remains strongly magnetized.
In summary, when the bubble expands non-relativistically,  it must be dissipative, while 
after breakout, the expansion becomes relativistic and the resistance falls so 
that the electromagnetic energy that is still being supplied by the source is mostly absorbed 
by the inflating bubble and by doing work against the surroundings.

\section{Optically thick expansion: mini-fireball}
\label{earlyT}

Under the  electromagnetic hypothesis, most of the energy released by 
the source comes out in 
the  form of Poynting flux. However  there must be some dissipation
that would lead to creation of  a 
lepto-photonic component (as discussed in \S \ref{dissipation}),
in case of non-relativistic stage
of bubble expansion inside a star  dissipation may be considerable). 
This will create an optically thick ''warm'' fireball (in a sense that it is dominated by magnetic field 
energy, but also has considerable  pressure). 
Expansion of this ''warm'' fireball will create a thermal precursor, similar to the
conventional FBM, but modified by presence of magnetic field \citep{lu00,lb03}.

At  early stages (before the breakout in case of long bursts)
the  plasma enthalpy is strongly dominated by 
lepto-photonic plasma with a temperature 
\be
T \sim \left( { L \over a \Delta \Omega r^2 \beta  c \Gamma^2 (1+\sigma)} \right)^{1/4}
\ee
where $\Delta \Omega$ is a typical opening  solid angle 
and luminosity of the source then can be written as
$
L= \int d \Omega \Gamma^2 r^2 \beta  c \left(b^{2}/2+ w \right)
\label{L}
$,
where $w$ is plasma enthalpy, $b$ is  a toroidal magnetic field in the plasma rest frame
times $\sqrt{4 \pi}$,

After breakout 
 the flow will accelerate to relativistic velocities.
Initially,  conical expansion is mostly
 pressure-driven, even in the strongly magnetized case
(in this case magnetic pressure gradients and hoop stresses balance out each other).
This results in  dynamics qualitatively similar to the unmagnetized case:
 the wind plasma accelerates $\Gamma \sim r$ while 
its density, pressure and temperature   decrease
$n\sim r^{-3}$, $p \sim r^{-4}$, $T \sim r^{-1}$,  so that 
magnetization parameter remains approximately constant  (\cite{lb03}).

When the  temperature falls below $\sim 10-20 keV$, most of the  pairs annihilate. 
This suddenly reduces the  optical depth to Thomson scattering below unity.
(Under certain conditions
photons may  remain trapped
in the flow. In this case, thermal
 driving  by photon pressure continues, until the  thermal photons
escape.)
As a result
the lepto-photonic part of the flow decouples from the  magnetic
field
and $\sigma$  increases by roughly seven  orders of magnitude
to $\sigma \sim 10^{9}$. 
 The
 thermal radiation from the lepto-photonic component
has a rest-frame temperature $T_0 \sim 10-20 keV$ times  a boost
due to the bulk motion.
This  thermal radiation, which should peak around $\sim 100 $ keV  may  put constraints on the
initial $\sigma$ \citep{lu00,dm02}. 

\section{Relativistic expansion}

\subsection{Short GRBs and long GRBs after breakout}

In case of long GRBs inflating a bubble inside a star, 
eventually the bubble will break free
out of the star forming two axial jets 
 along which 
Poynting flux will flow until  the central source slows down on the time scale $t_s\sim100$~s.
Outside the star, the bubble will expand ultra-relativistically and bi-conically.
For short GRBs, presumably associated with merger of neutron
stars in a low density environment,  there is no preceding non-relativistic
stage so expansion of the bubble is relativistic from the beginning.
 In case of relativistic motion
there is no necessity anymore to destroy magnetic flux through ohmic
dissipation:
 the effective load can
consist of the performance of work on the expanding blast wave. This is
where most of the power that is generated by the central magnetic
rotator ends up.

After the bubble  has expanded beyond a radius 
$
r_{{\rm sh}}\sim ct_s\sim3\times10^{12}t_{s,2}{\rm cm}
 $ ($\sim 10^{10}$ cm for short bursts)
the electromagnetic energy will be concentrated
within an expanding, electromagnetic shell with thickness $\sim r_{{\rm sh}}$
and with most of the return current completing along its
trailing surface (see Fig. \ref{current}). The global dynamics of this shell and its subsequent 
expansion are set in place by the electromagnetic 
conditions at the light cylinder and within the collimation region. 
An important property of ultra-relativistic outflows is that they are hard to collimate  
 \citep{Chiueh,bogo01}, 
so that any collimation should be achieved  close to the source, within a star,
where the flow is only 
 mildly relativistic. 

Interaction of the magnetic shell with 
the circumstellar medium proceeds in a similar way
to the non-relativistic expansion inside a star: 
the leading surface of the shell 
is  separated by a 
contact discontinuity (which actually becomes a rotational discontinuity
if the circumstellar medium is magnetized \citep{lyu02}). 
Outside the CD an ultra-relativistic shock front 
forms and propagate into the surrounding circumstellar medium.
The expansion is  non-spherical.
As long as the outflow is ultra-relativistic, the motion of the forward shock
is virtually ballistic \citep{sha79} and determined by the balance between the magnetic stress at the 
CD and the ram pressure of the circumstellar medium.

 A  type of collimation in case of electromagnetic explosions
is somewhat different from the conventional jet models of AGNs.
We expect that large Poynting fluxes
 associated with explosive release of $\sim 10^{51}$
ergs in case of long GRBs are sufficient to drive a relativistic outflow over a large solid angle,
so that during  the  relativistic stage the resulting cavity is almost spherical, but  the
Lorentz factor $\Gamma$ of the CD is a strong function of the polar angle.
The angular  distribution of magnetic field (and  of the Lorentz factor of the expansion)
depends on the dynamics of the bubble at the non-relativistic stage and the
distribution of the source luminosity.

In the framework of
electromagnetic model the outflow is strongly magnetized 
and  {\it subsonic}, 
$\sigma > \sigma_{crit}$, 
despite been strongly relativistic. 
In this case the  ejecta, in some sense,  may be considered as a collection of 
outgoing fast magnetosonic waves propagating
from the source to the contact discontinuity. Motion of the CD
is then  determined by the pressure balance between the Poynting flux from the source and
the ram pressure of the ISM. Thus, motion of the CD depends on the
source luminosity $L(t')$ at the retarded time $t'$ such that $R(t) = t -t'$.
In addition to forward flux, there is a  much weaker, by a factor
$\Gamma^2$, reflected flux
that propagates backward into the flow the information about the
circumstellar medium. Interference of forward and backward propagating waves
allows to define a finite Lorentz factor of the ejecta. 
The distribution of reflected current
is determined by the outgoing current and the boundary conditions.
At later times multiple reflections
from the contact discontinuity and the center become important as well.

\subsection{Stages of relativistic expansion of electromagnetic shell}

Relativistic 
 expansion of the  magnetized  shell may be separated into two stages, which we   will call
''early'' and ''late'', depending on whether or not most of the
fast waves emitted by the central source have caught up with the
CD and their energy has been given 
to the circumburst medium. The transition between two stages occurs  at the moment which
is similar to the deceleration
radius in fireball model, except that in the case of EMM the shell is decelerating 
all the time, but with different laws before and after the transition.
Keeping with the tradition we  will still call the transition radius as deceleration radius.

\subsubsection{ ``Early'' stage.}
\label{early}

At $r>r_{ph}$ the outflow becomes
a relativistically expanding shell of thickness 
$\sim ct_{{\rm s}}\sim3\times10^{12}\ {\rm cm}$  for long GRBs and
$ \sim3\times10^{10}{\rm cm}$  for short GRBs.
The shell  contains toroidal magnetic field;  the current now
detaches from the source and completes along the shell's inner surface.
At this stage the  CD  is constantly re-energized by the
fast-magnetosonic waves propagating from the central source.
The  average motion of the CD $R(t)$ is determined by the
average luminosity 
at the retarded time $t'$:
\be
L_\Omega(t') \sim   \rho c^3 \Gamma^4 R(t)^2 \beta^3
\ee
which for constant  luminosity gives
$\Gamma \sim (L_\Omega/\rho c^3)^{1/4}  r^{-1/2}$ (in a constant
density medium) or  $\Gamma \sim  (L_\Omega / 4  \kappa  \rho_0 r_0^2 c^3)^{1/4} = {\rm const}$
(in a $ \rho(r)=\rho_0(r_0/r)^2 $ wind).
If the central source releases most of the  current along the axis and the equatorial plane, as
  argued in \S \ref{Itheta}, then $\Gamma \propto 1/ \sqrt{\sin \theta}$ at this stage.
If source's luminosity varies, this will be reflected in the 
"jitter" of the CD. Development of instabilities at the 
CD, like 
impulsive  Kruskal-Schwarzschild instability \citep{lb03}
may lead to dissipation
and particle acceleration.
The internal structure of the magnetic shell is a messy
 mixture of the outgoing waves from the source and the ingoing
waves reflected from the CD, similar to  a pre-Sedov phase in
hydrodynamical explosions.
  Unlike the case of a
hydrodynamic  blast wave with energy supply, no internal discontinuities form
inside  magnetic shell.

Early stage lasts 
 for  $ct_{{\rm s}}<r<r_{{\rm dec}}$, where 
\be 
r_{{\rm dec}}=
(L_\Omega t_{{\rm s}}^2/\rho c)^{1/4} 
\sim 3 \times 10^{16}
L_{50}^{1/4} t_{s2}^{1/2} n^{-1/4}{\rm cm},
\label{rsh}
\ee  for long bursts
(in the observer frame this phase lasts  $\sim t_s \sim 100$~s) 
and $r_{{\rm dec}} \sim 3 \times 10^{15} {\rm cm}$ for shorts (similarly, 
in the observer frame this phase lasts  $\sim t_s \sim 1$~s).
 Radius $r_{{\rm dec}}$ (\ref{rsh}) is somewhat similar to the deceleration radius in case
of FBM; at this moment most energy of the shell in given to the circumburst medium,
$L_\theta t_s \sim 
E(\theta) \sim \rho  c^2 r_{{\rm dec}}^3 \Gamma(r_{{\rm dec}},\theta)^2$ (note that here 
$ \Gamma=\Gamma(r_{{\rm dec}})$, not $\Gamma_0$ as in case of FBM, 
since there is no formal definition of $\Gamma_0$ in case of EMM).

\subsubsection{''Late stage''  ($r_{{\rm dec}} <r<r_{{\rm NR}}\equiv
(L_\Omega t_{{\rm s}}/\rho c^2)^{1/3}$).}

At distances $r >r_ {{\rm dec}}$ most of the
 waves reflected from the CD have propagated throughout the shell, so that  all the regions of the shell come
into  causal contact.
Most of the energy of the explosion will reside in the 
blast wave which will eventually settle down to follow a self-similar expansion.
As the expanding shell performs
 work on the surrounding medium its total energy
decreases;
the amount of energy that remains in the ejecta shell during  the late
stage is small, $\sim E_\Omega /\Gamma^2$.
 Most of the energy  is still
concentrated in a thin
shell with $\Delta R  \sim R/ \Gamma^2$ near the surface of the shell
which is moving according to $\Gamma \sim \sqrt{E_\Omega / \rho c^2}\, r^{-3/2}$
(in a constant density medium), or $\Gamma \sim r^{-1/2} $
 (in a $\rho \sim r^{-2}$ wind).
If the central source releases most of the  current along the axis and the equatorial plane, as
   discussed in \S \ref{Itheta}, then $\Gamma \propto 1/\sin \theta$ at this stage.
[Note that at the ''early stage'' $\Gamma \propto 1/\sqrt{\sin \theta}$, but no lateral
 re-distribution of energy is required at the transition
 since transition between ''early'' and ''late''
stages occur at different times for different $\theta$.]
 The energy of the shell decreases only weakly
with radius, $dE/dt \sim 1/r $ in constant density and
 $dE/dt \sim 1/r^3$ in a wind, so that the surface of the shell keeps moving
relativistically as long as the preceding shock wave is moving relativistically,
until $r \sim (E_\Omega /\rho c^2)^{1/3} \sim 10^{18}$ cm - the shock never
becomes completely free of the shell  \citep{lb03}.
Interestingly, the   structure of the {\it magnetic}
 shell (in particular the  distribution of energy)
 resembles at this stage the structure of the {\it hydrodynamical}
 relativistic blast wave  \citep{blm76}. 
This can be formally understood by noting
that for motion perpendicular to magnetic field dynamical equations for magnetized flow can be reduce
to non-magnetic case, but with a different equation of state \citep{LLVII}.

''Late stage'' of magnetic shell expansion 
corresponds to  the conventional  afterglow phase when synchrotron and inverse Compton radiation 
is emitted throughout the electromagnetic spectrum. The initially 
aspheric expansion  will give the appearance of a jet with the ``achromatic break'' 
occurring when the fastest  Lorentz factor of the spine $\Gamma(\theta=0)$
  becomes comparable with the reciprocal
of the observer's inclination angle with respect to the symmetry axis, 
$\Gamma(\theta=0) \sim 1/\theta_{ob} $.
When $r> r_{{\rm NR}}\sim(Lt/\rho c^2)^{1/3}\sim2 \times 10^{18}L_{50}^{1/3}t_{s2}^{1/3}
n^{-1/3}$~cm, the blast wave become non-relativistic and will become more 
spherically symmetric, while evolving towards a Sedov solution.

\section{Production of GRB}

By the time the shell radius expands to  $r_{\rm dec}$
most of the 
electromagnetic Poynting flux from the source will have  caught up with the CD
and been reflected by it, transferring its momentum to the blast wave.  
Simultaneously a strong region of magnetic shear is likely to develop at the {\it  outer } 
part of the
CD \citep{lyu02}. 

We  propose that the {\it  $\gamma$-ray-emitting electrons are accelerated
near $r_{\rm dec} \sim 10^{15}- 10^{16} $ cm (for long bursts) and $\sim  10^{15}$ cm (for short bursts)
 due to development of electromagnetic current-driven  instabilities }
(conventional model of particle acceleration  -
 acceleration at internal shocks - cannot work in
 this model since in the limit $\sigma \gg 1$  fast shocks are either weak or do not  form
 at all).
 The development of
 current instabilities usually results in enhanced  or anomalous
 plasma resistivity  which leads  to an efficient
 dissipation of the  magnetic field.
 The magnetic energy is  converted into heat, plasma bulk motion
 and, most importantly, into   high energy particles, which, in turn,
 are responsible for the
   production of the
   prompt $\gamma$-ray emission.  
    The conversion of magnetic energy
    into particles may be very efficient.
    For example, recent RHESSI observations of the Sun indicate that, in reconnection
    regions, most magnetic energy goes into non-thermal electrons
    \citep[][]{benz03}.

To illustrate how magnetic dissipation may proceed, we  describe shortly
physics underlying the  development of the so-called tearing mode.
Consider a smooth distribution of electrical current, which can be viewed as
a set of many small current wires. Since parallel currents attract, 
such a system is likely to develop narrow current sublayers where
dissipation, which is inversely proportional to the square of magnetic field gradient,
becomes high. In addition, high anomalous resistivity, proportional to local 
current density is likely to develop. Similar to non-relativistic plasmas,
in strongly magnetized plasmas
tearing mode develops on time
scales {\it much shorter than resistive time scale in the bulk}
\citep{l03}. The final outcome of the development of the 
tearing mode is formation of reconnection sites
and dissipation of magnetic energy.

Particle acceleration  by dissipative magnetic fields
 may proceed in a number of ways. The best studied non-relativistic
 example is particle acceleration in reconnection regions either
   by inductive
   electric fields outside the current sheet or  resistive electric fields inside the
    current sheets \cite[\eg][]{cl02} or formation of shocks in the downstream
    of reconnection regions \citep[\eg][]{bf94}. Investigation of the particle
    acceleration in the relativistic regime of reconnection is only beginning \cite[\eg][]{llm02}.
    Relativistic reconnection may  produce power-law spectra of accelerated particles
    \citep[][]{Hosh,llm02}.
    For example, in the relativistic Sweet-Parker reconnection model
    \citep{lu03},
    if one balances linear acceleration inside the reconnection layer by the
    resistive electric field, $d_t  {\cal E} \sim e E c$ with the rate
    of particle escape (proportional to relativistic gyro-frequency),
    $d_t \ln N({\cal E}) \sim \om_B (mc^2 / {\cal E})$, one finds
    $
    N({\cal E})  \sim {\cal E}^{-\beta_{in}}
    $
    where ${\cal E}$ is the energy of a particle, $N({\cal E})$ is the particle number
    and $\beta_{in}  $ is the inflow velocity   \citep{Hosh}. For relativistic reconnection
    the inflow velocity can be relativistic \citep{lu03},
$\beta_{in} \rightarrow 1$. 
    The fact that reconnection models can produce spectra
    which are prohibitively hard for shock acceleration may serve as a distinctive property
    of electromagnetic models.

In addition to the acceleration mechanisms which are based on known non-relativistic
schemes, it is feasible that acceleration in relativistic, strongly magnetized plasma
may proceed through mechanisms that do not have non-relativistic or fluid analogues.
Examples of this type of acceleration include 
particle acceleration 
    through
     formation of a spectral cascade of nonlinear  waves in force-free plasma which
     transfer energy to progressively larger wave vectors until this energy
     is taken up in accelerating a population of relativistic electrons and
     positrons \cite{tb98}. 
     Since for $\sigma > 1$ cascade is likely to be terminated at plasma frequency, 
which is lower by a factor $\sqrt{\sigma}$ the cyclotron frequency, the likely
emission mechanism in this case is inverse Compton scattering.
Another possibility is development of 
kinetic electromagnetic-type
  instabilities of the shell surface currents, as proposed  by
  \cite{su96} and in somewhat different form by \cite{Liang}.
  Since studies of kinetic properties of strongly magnetized relativistic plasmas are
  only beginning, it is hard to predict acceleration efficiency and particle spectra.
  Numerical studies in the coming years will be most important here.

\section{Production of afterglows}
\label{after}

Except at early stage (as discussed in \S \ref{early}),
afterglows are generated in a similar way both in the FBM and EMM.
As the magnetic  shell expands, its energy is gradually transfered
 to the preceding forward shock wave. 
Relativistic particles are accelerated  in the blast wave producing
the observed  afterglow in a manner
which is similar to what is proposed for fluid models,
except that contact discontinuity itself may be an  important source
of magnetic flux  through impulsive  Kruskal-Schwarzschild instability
\citep{lb03}, 
 so that  afterglows may  result from a mixture
of relativistic particles, derived from the shock  with 
magnetic field derived from the shell. 

At late times, well 
beyond $r_{\rm dec}$ (which in observer frame is  nearly  coincident
   with the prompt phase),
   the 
  temporal behavior of proper afterglow (as opposed to tails of prompt emission, see
  \S \ref{Emissionradius}) 
  is  determined by the total energy release $E_\Omega $ 
  and not by the form
   of that energy.  As a consequence, late afterglow observations
   can hardly be used to distinguish between the models. 
   The only property of the source that the forward shock
   ''remembers'' at late times is the angular  distribution of the
   deposited energy $E(\theta)$
   (there is little sideways evolution in relativistic regime \citep{sha79}).
 Thus,
 the angular  distribution of the total energy $E(\theta)$ can
 be used to distinguish between different models, if a model predicts it.
  
 In case of EMM, the preferred  lateral distribution of the   magnetic field, 
  energy fluxes and luminosity correspond to the  line current, \S \ref{Itheta},  so that
$L \sim 1/\sin^2 \theta$.
 This 
  translates into distribution of Lorentz factors of the forward shock
\be
\Gamma \sim  \left( {  E \over  \rho_{ex} c^5} \right)^{1/2} { t^{-3/2} \over 
\sqrt{ \theta^2+  \theta_0^2}} 
\ee
where $\theta_0$ is the angular size of the core of the jet.
(Its  minimal
 size is 
magnetic Debye radius, $r_D=\sqrt{ I \over  2 \pi n e c}$, which gives $
\theta_0 \sim \left( { m^2 c^5 \sigma^2 \Gamma^2 \over L e^2}  \right)^{1/4}
\approx 10^{-3} L_{50}^{-1/4}  \sigma_9^{1/2}\Gamma_2^{1/2}$,
\citep{lb03}.)
This type of shock has been named ``structured jet''
(or universal jet) \citep{Lipunov,ros02}, though in our
model there is no proper ``jet'', but simply a  non-spherical outflow. 
The most intense bursts and afterglows
in a flux-limited sample will be seen pole-on
and can exhibit achromatic breaks when $\Gamma\sim\theta^{-1}$,
which might be mistaken for jets.
 
In conclusion,  observational appearance of  GRB afterglows
 depends mostly on two
parameters: (i) explosion energy 
(more precisely, on the
 ratio on the explosion energy to circumstellar density) and
 (ii) the
viewing angle that the progenitor's axis is making with the line of sight.
This possibility, that all GRBs (and XRFs) are virtually 
the same but viewed at different angles
resembles  unification scheme of AGNs.

\section{Tests of GRB models}

\subsection{Testing  the fireball model: reverse shock emission}

Perhaps the simplest test of GRB models could have come
 from observations of emission from the  reverse
shock propagating in the ejecta,  which typically falls
into the optical range. 
FBM predicts 
strong reverse shock emission, so that
absence of nearly contemporaneous optical emission in most GRBs would be 
a strong argument against
FBM. In MHD models with $\sigma > 1$ reverse shock is weak, while 
 in EMM reverse shock  is absent altogether. 

Since possible observation of reverse shock emission may play a decisive role 
in validating a GRB theory, 
next we  discuss briefly properties of the reverse shock expected 
{\it within a framework of FBM}
\cite[for more extensive discussion see, \eg][]{sari99,kobayashi}.
In the framework of FBM both reverse shock and  internal shocks which produce
prompt $\gamma$-ray emission originate in the same fluid and have similar properties 
(\eg being weakly relativistic). 
One can naturally expect that 
microphysical properties of particle accelerating, being complicated
and poorly understood, are the same for the same type of shocks.
Thus, the conventionally introduced quantities like $\epsilon_B$
(magnetic field value with respect to equipartition), $\epsilon_e$ (electron energy density  with respect to equipartition),
 and $\gamma_{min}$ (minimum Lorentz factor of accelerated electrons) {\it  must be the
same for both cases}.
As a result, for any given burst, observations of the prompt emission  can be used to 
predict properties of a corresponding optical flash. 

The amount of energy dissipated in reverse shock is comparable
to the energy dissipated in the forward shock and to the total GRB energy \citep{sari99}.
The principal difference between prompt and optical  emitting electrons is 
the radii of emission and ratio of cooling to expansion time scales.
In a framework of FBM, prompt emission is generated at distances
  $r_{GRB}\sim 2 \Gamma_0^2 \delta t c \sim 10^{12}-10^{13}$ cm 
($\delta t \sim 0.01$ s is variability time scale
 of the central source and, 
 within a framework of FBM, of the prompt emission),
 while 
reverse shock emission is typically generated at distances $R \sim 2 c t_s \Gamma_0^ 2 \sim  
10^{16} $ cm
(seen at observer time
$t_{obs} \sim  t_s$; we  concentrate on a simple so called ''thick shell'' case).
Using conventional fireball parameterization for minimum
Lorentz factor of accelerated particles $\gamma_{min} \sim \epsilon_e (m_p/m_e) \Gamma_s$ where $\Gamma_s$ is the shock Lorentz factor,
and parameterizing energy density of magnetic field in the plasma rest frame to ion energy density,
$\rho c^2 =L/(4 \pi \Gamma_0^2 r^2 c)$ and
$ b = \sqrt{\epsilon_B} \sqrt{ 8 \pi \rho c^2}$,
 the ratio 
  of prompt to reverse shock
 frequencies is
 \be
 {\om_{GRB} \over \om_{RS}}\sim
 { r_{RS} \over r_{GRB} } \left( { \Delta \Gamma \over \Gamma_{RS}} \right)^2 \sim { r_{RS} \over r_{GRB} }  \sim{t_s \over \delta t}
 \ee
 so that 
the 
peak of reverse shock emission occurs at  $\sim 1-10$ eV.

An important qualitative difference between prompt and optical emitting electrons 
is that the former are in fast cooling regime, while the latter  are in slow cooling regime.
The radius beyond which optically emitting electrons enter slow cooling regime,
\be
r_{cool} \sim \epsilon_B \epsilon_e { (\Gamma_{RS}-1) L m_p \sigma_T \over 3 \pi c^3 m_e^2
\Gamma_0^3} \sim 2 \times 10^{14} \, L_{50} {\rm cm},
\ee
($\Gamma_{RS}-1\sim 1$ and $\Gamma_0 =300$ was assumed)  is typically smaller that
$R_{RS}$. As result, only small fraction of energy received by an optical electron
$\sim R_{RS} /(\Gamma_0 r_{cool}) \sim 0.02$ is emitted; the rest is lost to adiabatic
expansion. 
For a GRB  of $E_\gamma \sim 10^{51} $ ergs, optical flash would have  $E\sim 10^{49} $
 ergs.
For a typical GRB  burst with fluency $\sim 10^{-6}$ erg/cm$^2$, and duration
of  $\sim 100$ sec, this will result in optical flash of  magnitude $\sim 12m$.
Even if we increase the estimate of   $R_{RS}$ and duration of optical flash  each
by an order of magnitude, resulting  optical flash would have $\sim 17m$.
On the other hand, brightest bursts may
 reach fluences $\sim 10^{-4}$ erg/cm$^2$, which, according to these estimates, 
can produce optical flash of $\sim 7m$.
    In addition, adiabatic cooling results in flux decay $\propto t^{-2}$ and 
    a clear radio signal is expected \citep[\eg][]{nakar}.
Thus, fireball model makes a
 predictions that all GRBs must have optical
     flashes in the range $12m-17m$, with some variations of
         few magnitudes  (both brighter and dimmer) depending on particular properties
	     of each burst.

In the Swift era, {\it not a single GRB has shown the predicted behavior}.
This is despite the fast on-board optical telescope and a large number of 
ground based robotic telescopes (RAPTOR, ROTSEE, TAROT and others).
{\it
Reverse shock emission is virtually an  unavoidable
 prediction of the fireball model, so that  
 absence of predicted  reverse shocks  emission in the Swift era
 argues against the fireball model.}
Naturally, there is a number of ways that
 through which  optical flashes can be suppressed
(\eg
cooling of optically emitting electrons on photons of prompt emission \cite{belob},
''thin
shell case", 
when the  reverse shock emission is spread over longer times, producing
weaker signal \cite[\eg][]{mcmahon}).
 {\it  A possible 
 explanation of an  absence of clear reverse shock signal is that  
ejecta plasma is strongly magnetized. } In the case when energy density of magnetic field
dominates  the total energy density  ($\sigma \geq 1$) 
reverse shock becomes very inefficient in dissipating flow energy \citep{KC84}.

Irregular  
optical flashes (like  GRB 050525a \citep{klotz} and  GRB 050904 
\citep{boer}) may be produced by other mechanisms, like  
gamma-ray pair production in front of the forward shock, \cite{belob,tm00},  or
be a low energy tail of prompt emission.

\subsection{Electromagnetic model: bright early afterglows}
\label{spectr}

\begin{figure}
\includegraphics[width=0.95\linewidth]{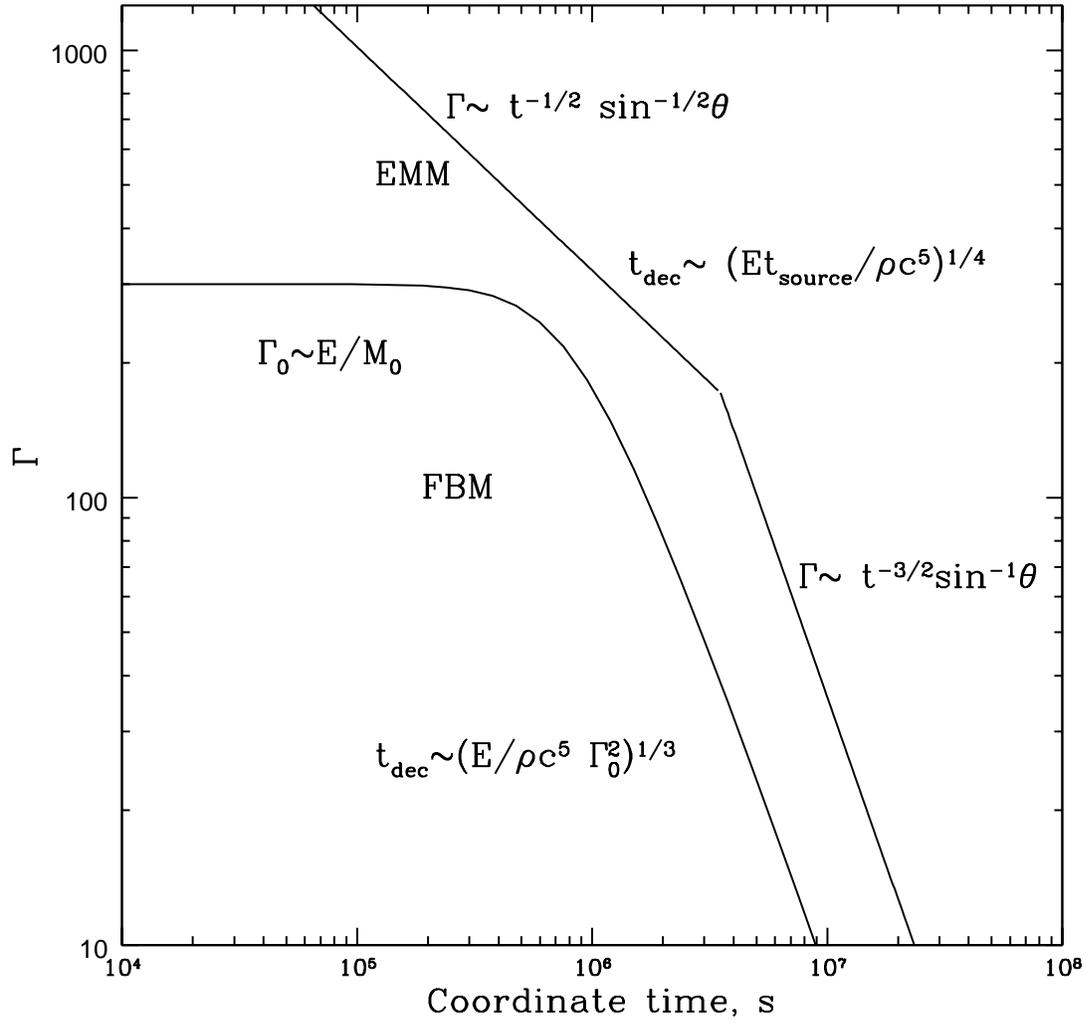} 
\caption{Evolution of Lorentz factors in the fireball  and electromagnetic models
 for constant external density.
Relative normalization of the curves depends
on the viewing angle (in the EMM).}
\label{fig}
\end{figure}

\begin{figure}
\includegraphics[width=0.95\linewidth]{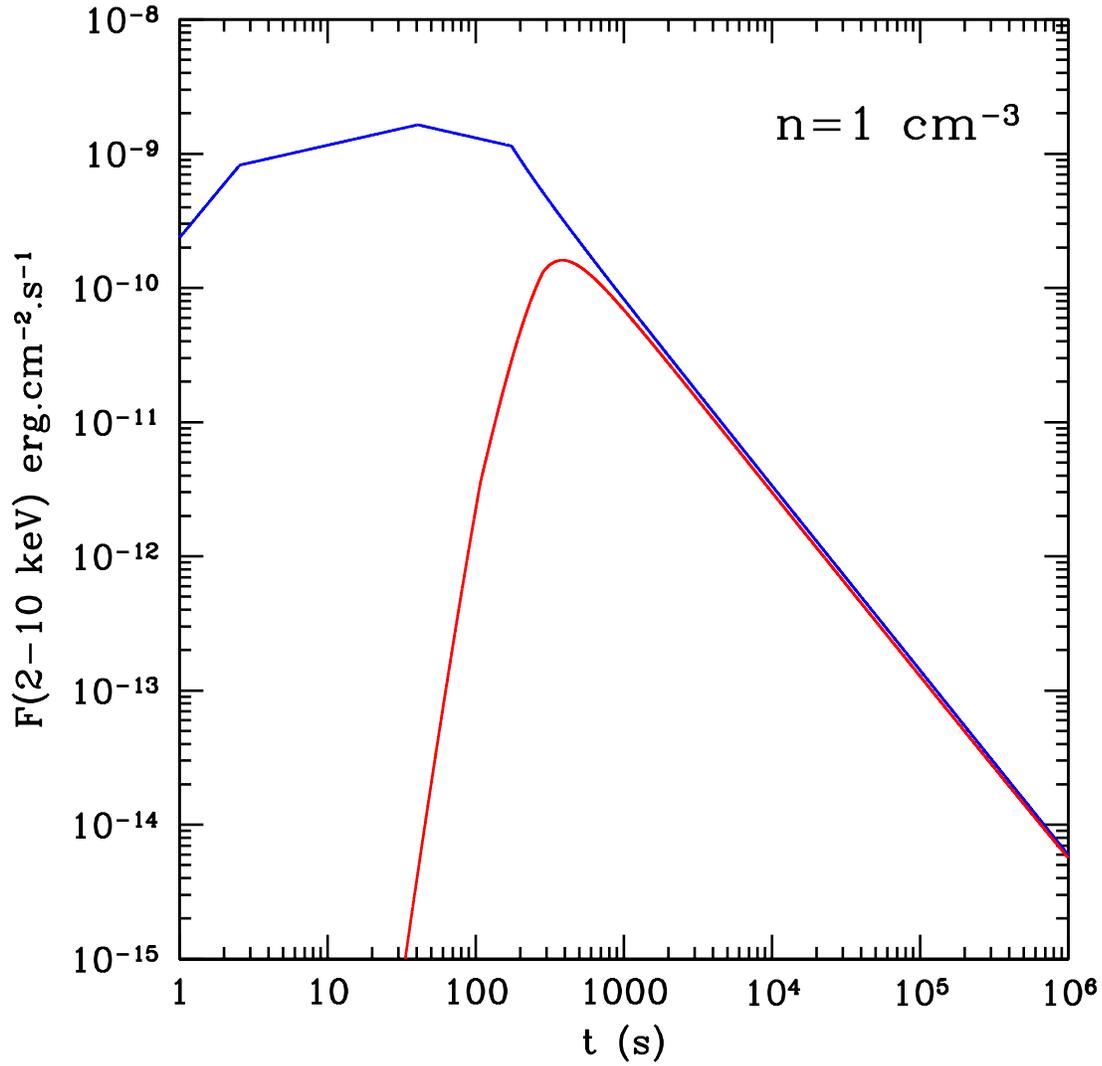}
\caption{X-ray afterglow in the 2-10 keV energy band for a uniform medium of
density $n=1$ cm$^{-3}$ and injected power $L_{iso}=10^{52}$ erg s$-1$. 
 The source is active for 50 s,
the assumed redshift is z=2.5, fraction of energy in electrons $\epsilon_e=0.1$,
fraction of energy in magnetic field $\epsilon_B=0.001$.  Blue line: electromagnetic model, 
red line: fireball model (from Mochkovitch \etal, in prep.)}
\label{mochkovXray}
\end{figure}

Fireball and electromagnetic models 
make very different prediction for the properties of early afterglows
\citep[see Fig. \ref{fig},][]{lyut04}.
According to EMM, at the early afterglow stage the Lorentz factor and peak frequency are
 larger (and falling with time) than in the
FBM (constant  Lorentz factor and  peak frequency).  {\it  Early afterglow in the EMM
are more energetic than in FBM}, Fig. \ref{mochkovXray},
and can blend with the prompt phase.

\subsection{Emission radius of prompt photons and early Swift afterglows}
\label{Emissionradius}

One of the surprising early results from Swift satellite was detection of X-ray spikes
and breaks in light curves at intermediate times,
much longer than burst duration but well before the conventional 
jet break \cite[\eg][]{Tagliaferri,Nousek,Chincarini}. 
A typical behavior includes fast-slow-fast decay with transitions near $100-1000$ seconds 
and $\sim 10^4$ seconds.
This presents a real  challenge to GRB models, since if the emission is seen
''head on'', within angle $\theta \leq 1/\Gamma$, the 
radii at which these  features should be  produced correspond 
to radii much larger than deceleration radius. At these times
 most of the energy is in the forward shock which  should 
produce smooth light curve. (Late time injection or specific distribution of Lorentz
factors are some possibilities discussed \citep[\eg][]{Zhang,Lazzati05}. For discussion
in the framework of the cannonball model see \cite{Dar05}.) 

The initial fast decaying part of afterglows was argued to be a ''sideways'' prompt emission,
coming from angles $\theta > 1/\Gamma$ \citep{Kumar00,Barthelmy05}. 
If this interpretation is correct, one can 
determine emission radii of the prompt emission and compare them with model predictions.
(We remind that FBM predicts radii of emission $r_{em} \sim
 2 \Gamma_0^2 c \delta t \sim 10^{12}-10^{13}$
cm,
  while EMM predicts  $r_{em}\leq r_{dec} \sim  10^{16}$ cm, Eq. \ref{rsh}). 
If emission is generated at $r_{em}$ and is coming to observer 
 from large angles,  $\theta > 1/\Gamma$, its delay with respect to the start of the  prompt pulse
is $\Delta t \sim (r_{em}/c) \theta^2/2$. For typical observer angle $\theta \sim 0.1$ and 
first break of a light curve at $\Delta t \sim 1000$ seconds, the implied emission radius is
$r_{em}\sim  6 \times  10^{15}$ cm. {\it This is at least two orders of magnitude  larger
than is assumed in the fireball model, but is close to the assumption of the electromagnetic model.}
[To be consistent with FBM and variability on short times scales, the
Lorentz factor of the flow should be $\Gamma_0 \sim 3000$, but this  would imply
that emission is strongly de-boosted, $\Gamma_0 \theta \sim 300 \gg 1 $.]
{\it Interpretation of light curves breaks at $\sim 10^3$ s as 
been due to prompt emission seen at large angles, $\theta > 1/\Gamma$, is inconsistent with
the fireball model.}

\subsection{Fast variability from large radii}

If prompt emission is produced at distances $\sim 10^{15}-10^{16}$ cm,
how can fast variability,
on times scales as short as milliseconds, be 
achieved? One possibility, is that emission is beamed in the outflow frame, for example 
due to 
 relativistic motion of ''fundamental emitters'' \citep{lb03}.
Possible origin of relativistic motion of ''fundamental emitters'' may be the fact that in case
of relativistic reconnection occurring in plasma  with $\sigma \gg 1$, the outflowing matter reaches
relativistic speeds with $\gamma_{out} \sim \sigma$ \cite[]{lu03}.
Internal synchrotron emission by such jets,
or Compton scattering of ambient photons will then be strongly beamed in the frame of the outflow.

\begin{figure}
\includegraphics[width=0.95\linewidth]{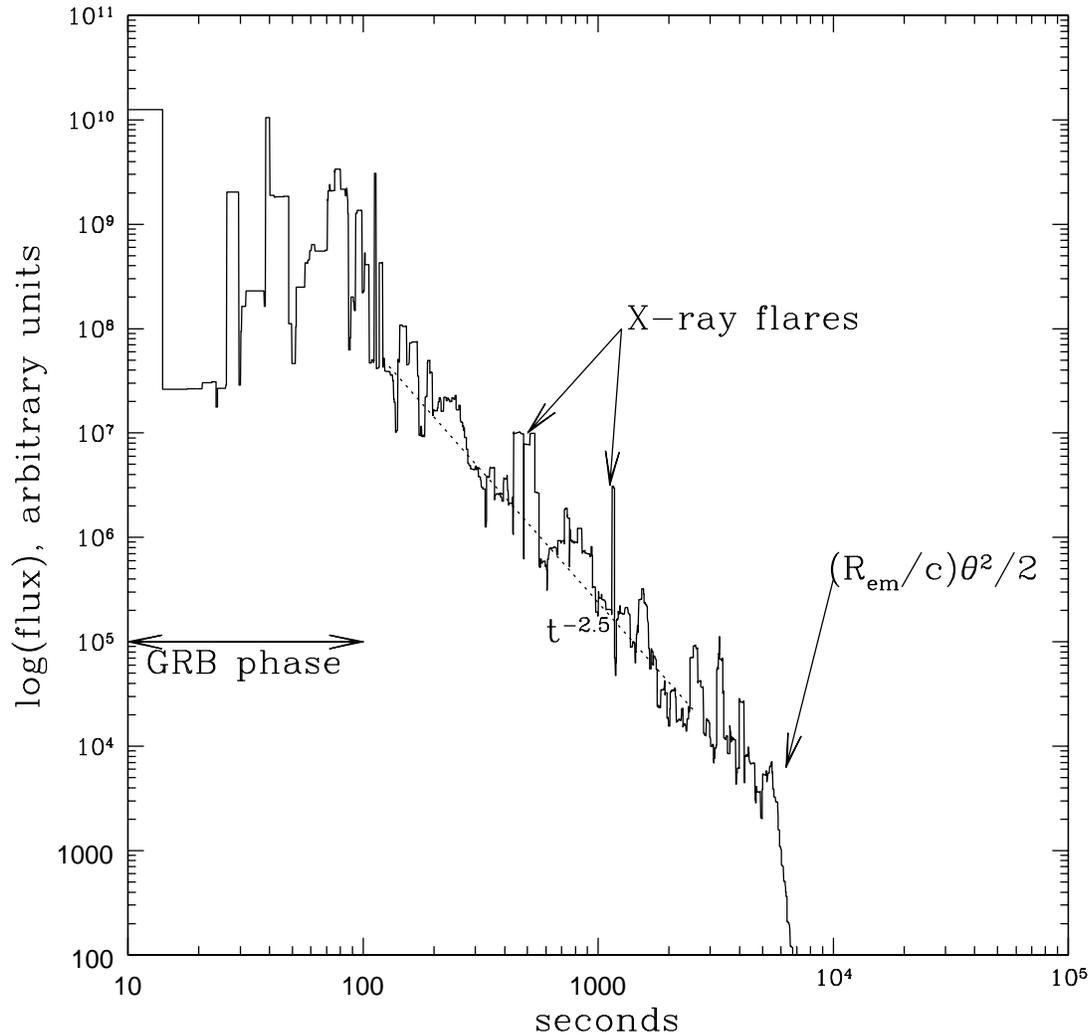}
\caption{Prompt emission produced by emitters  moving  randomly  in the bulk frame.
Emission is generated within a shell of thickness $t_s c = 3 \times 10^{12}$ cm 
moving with $\Gamma =100$  at distance $r_{em} = \Gamma^2 t_s c $ by randomly distributed
jets with random orientation moving with random Lorentz factors $1<\gamma_T < \gamma_{T,max} =5 $.
Each emitter is active for random time $0<t'_{em} < 0.5 t_s c \Gamma = t_{pulse,max}$ 
in its rest frame.
Homogeneous jet centered on an observer with opening  angle $\theta =0.1$,
dimensionless parameters ${N \pi /(\Gamma \gamma_{T,max} \theta)^2} = 1.2$ (probability of seeing 
one sub-jet ''head-on'' from angles $< 1/\Gamma$) and
${N (c t_{pulse,max}/2)^2 / r_{em}^2 \theta ^2 t_s c \Gamma} =0.19$ (efficiency of
energy conversion), where $N$ is total number 
of emitters. Intensity of  emission is $\propto \delta^{3+\alpha}$,
where $\delta $ is Doppler factor and $\alpha=0.5$ is spectral index.
As the burst progresses, the average Doppler factor $\delta \approx 
 t_s \Gamma / t $ and the average flux decays as $t^{-(2+\alpha)}= t^{-2.5}$
 in accordance with analytical estimates \citep{Fenimore}
 (Lyutikov, in prog.)
}
\label{GRBafter}
\end{figure}

Consider an outflow moving with a bulk Lorentz factor $\Gamma$ 
with randomly distributed emitters moving  with respect to the shell rest frame with
 a typical Lorentz factor $\gamma_T$. Highly boosted emitters, moving towards an  observer,
have  Lorentz factor $
 \gamma \sim 2 \gamma_T \Gamma
$, so that   
  modest values of $\gamma_T \sim 5-10 \ll \Gamma \sim 100-300 $
 suffice to produce short time scale  variability  from
  large distances.
As the burst progresses, larger angles and more of internal
 jets producing prompt emission become visible. Most of them will be seen from
large angles  $> 1/\gamma_T$ in the bulk frame, producing smooth curves.
Occasionally, a jet at large viewing  angle, $\theta > 1/\Gamma$, but
 directed towards an observer will be seen, producing an X-ray flare.
One expects a break in the light curve at $\Delta t \sim (r_{em}/c) \theta^2/2$, where
$\theta$ is a viewing angle (in a structured jet model, this is the angle between the
jet axis and direction to the observer).
Afterglow should start to blend with prompt emission at later times.
In Fig. \ref{GRBafter} we  plot an example of a prompt light curve 
in this model (Lyutikov, in prog.). The model readily explains many unusual properties
of early afterglows:
(i)  X-ray flares and light curve breaks  at late times, 
much longer than conventional prompt  GRB
 duration (extended source activity is not needed!),
(ii) fast variability,
 (iii) gradual softening of the spectrum, (iv) hardernig of a  spectrum during  X-ray flares 
\citep{Burrows}.

\subsection{Observational implications of the electromagnetic model}
\label{implic}

In this section we  give a short discussion of how the main GRB phenomena are  (or may be) explained
 within a framework of EMM.
\begin{itemize}
\item {\bf Jet break in afterglow}
GRB outflows have large opening angles, but do not have a jet in a proper sense.
Outflows are non-isotropic so an achromatic break is inferred when the viewing
angle is $\theta_{ob} \sim 1/\Gamma$. 

\item 
{\bf Structured jet}. 
 The model predicts and  gives a theoretical foundation for  the  ``structured jet''
profile of the external shock. 

\item
{\bf XRF flashes}.
Another testable prediction of the model is that
much more numerous X-ray flashes (XRFs) should be observed, which  may be coming
``from the sides'' of the expanding shell, where the
 flow is less energetic and the
Lorentz boosting is weaker. In addition, the total {\it  bolometric } energy 
inferred for XRFs  (from observations of afterglows before
radiative losses become important) should be comparable to the total bolometric energy
of $\gamma$-ray bursts.
Generally, the  distributions of 
parameters of XRFs should continuously match those of GRBs. 

\item {\bf Weak thermal precursor.} If a fraction $1/\sigma\sim 0.01-0.1$
of the magnetic energy is dissipated near the source, this should produce
a thermal precursor with luminosity $\sim 0.01-0.1$ of the
main GRB burst. 

\item
{\bf Hard-soft evolution }.
The trend of  GRB spectra to evolve from hard to soft during  a pulse
is  explained as a   synchrotron radiation in an expanding  flow with  magnetic field 
decreasing with radius $B \propto \sqrt{L} /r$ (later in a pulse emission is produced
further out where magnetic field is weaker, so that the peak energy will be lower;
 this is similar to "radius-to-frequency mapping" in radio
pulsars and AGNs).

\item  {\bf Amati $  E_{peak} - L$ correlation}.
A correlation between peak energy and 
total luminosity, $E_{peak} \sim \sqrt{L}$ \citep{Amati}
 follows from the assumption of a  fixed typical
emission radii and fixed minimum particle energy since
$B \sim \sqrt{L_\Omega}$.

\item  {\bf Variability}.
 Variability of the prompt emission reflects 
the statistical properties of dissipation
(and not the source activity as in the FBM). 
Magnetic fields are non-linear dissipative  dynamical system which often show 
bursty behavior with 
power law  PDF.
[For example,
solar flares  show variability on a wide
range of temporal scales, down to minutes, 
which are  unrelated to the time scale of 22 years of magnetic field generation in the tachocline.]

\item  {\bf Prompt and afterglow polarization}.
Claims of high polarization \citep{coburn03,Willis} 
if confirmed, 
may provide a decisive test of GRB models  \citep[see though][]{rutl03}.
The best way to 
produce polarization in the range $10\% \leq \Pi \leq 60\%$
is through synchrotron emission in  large scale magnetic fields \citep{lyu03c}.
(Larger polarization can only be produced with inverse Compton mechanism, 
smaller  polarization can be produced by small scale magnetic fields.)

Large scale field structure in the ejecta 
  emission may also be related to  polarization 
of afterglows if fields from the 
 magnetic shell  are mixed in with the shocked
circumstellar material. 
In this case,
{\it the position angle should 
not change through the afterglow} while 
if polarization is observed both in prompt and afterglow emission 
the position angle should be the same. 
Also, polarization should be most independent of   the "jet break" moment. 

\end{itemize}

\section{Conclusion}

In this contribution I outlined the underlying assumptions for
  the 
 ``electromagnetic hypothesis'' for ultra-relativistic GRB
outflows.
 The most striking implications of the electromagnetic
 hypothesis is 
 that particle acceleration in the sources is due to direct dissipation of 
electromagnetic energy
rather than shocks and that the outflows are cold,
electromagnetically dominated flows, 
at least until they become strongly dissipative.

One of the major drawback of the model is that magnetic dissipation and particle acceleration
are very complicated processes, depending crucially on the kinetic and
geometric properties of the plasma. This situation may be contrasted with the shock acceleration
schemes, where a qualitatively correct result for the spectrum of accelerated
particles, {\it a kinetic property}, 
can be obtained from simple {\it  macroscopic } considerations (jump conditions).
Example of the Solar corona shows that despite being complicated magnetic dissipation 
is an effective mean of particle acceleration.

I have discussed possible observational 
tests of the hypothesis. In particular, 
interpretation of early afterglow features as being  due to 
prompt emission seen at large angles, $\theta \geq 1/\Gamma$, 
allows to  measure radius at which prompt emission
has been produced. Large prompt emission radii, $\sim 6 \times 10^{15}$ cm
 seem to be inconsistent with 
the fireball model, but close to prediction of the electromagnetic model.
Internal relativistic motion of ''fundamental emitters" assumed within  
EMM  may also explain X-ray flares during early afterglow phases
(without a need for long source activity).
An important implication of the electromagnetic model is that supernova
explosions may be magnetically driven as well \citep{lw71,bk71,Wheeler,prog03}. 

Over the years I have benefited from discussion with many colleagues, too numerous to be named here.
In preparing this contribution I am grateful to Roger Blandford, Tomas Janka, Davide Lazzati,
Ehud Nakar, Maurice van Putten  and  Stephan Rosswog
for  discussions and comments. I am also indebted to Robert Mochkovitch 
for shearing  his unpublished results.

\appendix

\section{Applicability of fluid approach for blast wave}

 In case of extremely high 
Lorentz factors of the ejecta (which require even higher values of 
$\sigma$ than were assumed in this paper),
fluid approximation for interaction of magnetized ejects
with ISM may break down. Consider an interface between ejecta and the surrounding medium
in its rest frame. As a particle from the
surrounding medium enters ejecta, it starts gyrating in  magnetic field.
If a fraction $\sigma/(\sigma+1)$
of the source luminosity $L$ is in the form of magnetic field,  then the turn angle in rest frame in one dynamical time
is
 \be
 \om_B' t_{exp} \sim  \sqrt{ \sigma \over \sigma+1}
 {2  e  \sqrt{ \pi L}  \over c^{5/2} m_p  \Gamma^3}
 \ee
 In order to justify fluid approximation this should be larger than unity, which requires
 \be
 \Gamma \leq \left( { \sigma \over \sigma+1} \right)^{1/4}
 \left( { 4 \pi  e^2 L   \over c^5 m_p^2 }\right)^{1/6}
 \sim 4 \times 10^4
 \ee
 for $\sigma \geq 1$.
Thus, for any
  $\Gamma \leq 4 \times 10^4 $ an  ISM particle can complete a half turn on a time scale
  short if compared with the expansion time scale. In this case, in laboratory frame
   momentum of the ejecta will be given to the particles
   almost instantaneously.
For larger Lorentz factors the instantaneous hydrodynamical approximation
  is not applicable, but if particles are turned by an angle larger than $\sim 1/\Gamma$ (lager
       than $\sim \pi$  in the  observer frame) they will
             still be carried
	            with the flow. Since the
		             rest-frame magnetic field goes as $\sim 1/(t \Gamma(t))$,  approximately linearly
			              with time (for constant $\Gamma$),  the rotational phase
of  a particle increases only logarithmically,
          $\int \om_B' dt' \propto \ln t$. Thus, it takes a very long time for a particle
	            to complete one gyration and be expelled from the ejecta.
 In this case, the ejecta will be effectively loaded with ISM
particles. 

Finally, for very high Lorentz factors,
\be
 \Gamma \geq  \left( { \sigma \over \sigma+1} \right)^{1/4}
{ \sqrt{e} L^{1/4} \over c^{5/4} \sqrt{m_p}} \sim 8 \times 10^6
 \ee
    a particle  makes 
a turn of
less than $1/\Gamma$ (in the ejecta frame) on a dynamical times scale. In this case
the ejecta just passes through ISM without much interaction and without slowing
down.

\begin{thebibliography}{99}


\bibitem[Aloy \etal(2005)]{aloy05}
{{Aloy}, M.~A. and {Janka}, H.-T. and {M{\"u}ller}, E.},
 2005,
 {\aap},
 436,
 273

\bibitem[{Amati} \etal(2002)]{Amati}
Amati, L \etal, 2002, 
{\aap},
 390,
 81

\bibitem[Barthelmy \etal(2005)]{Barthelmy05} Barthelmy, S. D., \etal 2005,  astro-ph/0511576

\bibitem[Beloborodov(2002)]{belob} {{Beloborodov}, A.~M.},
2002, 
ApJ, 565,
808


\bibitem[Benz \etal(2003)]{benz03} {{Benz}, A.~O., {Saint-Hilaire}, P.}, 2003,
 {Advances in Space Research}, 
 32, 2415

\bibitem[Bisnovatyi-Kogan(1971)]{bk71}
{Bisnovatyi-Kogan}, G.~S., 1971, {Soviet Astronomy}, 14, 652

\bibitem[Blackman \& Field(1994)]{bf94}
{{Blackman}, E.~G., {Field}, G.~B.}, 1994, {Physical Review Letters}, 73, 3097


\bibitem[Blandford \& McKee(1976)]{blm76}Blandford, R. D. \& McKee, C. F. 1976 Phys. Fluids 19 1130


\bibitem[Blandford(2002)]{bla02}Blandford, R. D. 2002 Lighthouses 
of the Universe ed. R. Sunyaev Berlin:Springer-Verlag

\bibitem[{Bo{\"e}r}  \etal(2005)]{boer}
 {{Bo{\"e}r}, M. and {Atteia}, J.~L. and {Damerdji}, Y. and {Gendre}, B. and 
 	{Klotz}, A. and {Stratta}, G.},
	Submitted to Nature, astro-ph/0510381

\bibitem[Bogovalov(2001)]{bogo01}
{Bogovalov}, S., 2001, {\aap}, 371, 1155

\bibitem[Burrows \etal(2005)]{Burrows}
{Burrows}, D.~N. \etal, 2005, astro-ph/0511039 

\bibitem[Chincarini \etal(2005)]{Chincarini}
{Chincarini}, G. \etal, 2005, astro-ph/0511107

\bibitem[Covino \etal(2005)]
{Covi05a} Covino, S., Antonelli, L.A., Romano, P., et al. 2005a, GCN Circ. 3665  

\bibitem[Chiueh \etal(1991)]{Chiueh}
Chiueh, T., 
 Li, Z-Y. \& 
 Begelman, M., 1991, ApJ, 377, 462, 

\bibitem[Coburn \& Boggs(2003)]{coburn03} Coburn,~W., \& Boggs,~S.E.
2003, \nat, 423, 415

\bibitem[Craig \& Litvinenko(2002)]{cl02}
{{Craig}, I.~J.~D., {Litvinenko}, Y.~E.}, 2002, {\apj}, 570, 387 

\bibitem[Dado \etal(2005)]{Dar05}
{{Dado}, S. and {Dar}, A. and {De Rujula}, A.},  
astro-ph/0512196

\bibitem[De Villiers \etal(2005)]{Hawley}
{{De Villiers}, J.-P. and {Hawley}, J.~F. and {Krolik}, J.~H. and 
	{Hirose}, S.},
	2005,
	 {\apj},
	  620,
	  878



\bibitem[Daigne \& Mochkovitch(2002)]{dm02}Daigne, F.  \& {Mochkovitch}, R.,
2002, \mnras, 336, 1271 
 
\bibitem[Djorgovski \etal(1998)]{Djorgovski}
{{Djorgovski}, S.~G. and {Kulkarni}, S.~R. and {Bloom}, J.~S. and 
	{Goodrich}, R. and {Frail}, D.~A. and {Piro}, L. and {Palazzi}, E.
		},
		1998,
		{\apjl},
		508,
		17
 
 \bibitem[Drenkhahn  \& Spruit(2002)]{ds02}
 {{Drenkhahn}, G. and {Spruit}, H.~C.}, 2002,
 {\aap},
 391,
 1141


\bibitem[Fenimore \etal(1998)]{Fenimore}
{{Fenimore}, E.~E. and {Madras}, C.~D. and {Nayakshin}, S.},
1998, {\apj},  473, 998

\bibitem[Ferrari (2004)]{Ferrari}
{{Ferrari}, A.},
2004,
{\apss},
293,
15

\bibitem[Fox \etal(2005)]{fox}
  Fox, D. B, \etal, Nature in press, astro-ph/0510110

\bibitem[Ghirlanda \etal(2003)]{Ghirlanda03}
{{Ghirlanda}, G. and {Celotti}, A. and {Ghisellini}, G.},
2003, {\aap}, 406, 879

\bibitem[Gehrels \etal(2005)]{Gehr05} 
Gehrels,  N., Barbier, L.,  Barthelmy, S.D., et al. 2005, Nature 437, 851

\bibitem[Goldreich \& Julian(1970)]{gj70} 
{{Goldreich}, P., {Julian}, W.~H.}, 1970, {\apj}, 160, 971

\bibitem[Hawley \& Krolik(2005)]{Hawley05}
{Hawley}, J.~F. and {Krolik}, J.~H., 
2005, astro-ph/0512227 

\bibitem[Heyvaerts \& Norman(2003)]{HeyN03}
Heyvaerts, J. \&  Norman, C. 2003, accepted for publication in ApJ, astro-ph/0309132

\bibitem[Hjorth \etal(2003)]{Hjorth}
Hjorth, J. \etal, 2003, {\nat},  423, 847


\bibitem[Kennel \& Coroniti(1984)]{KC84}  Kennel, C. F., \& Coroniti, F. V. 1984a, ApJ, 283, 694

\bibitem[Klotz \etal(2005)]{klotz}
{Klotz}, A., 2005, {\aap},
439,
35

\bibitem[Kobayashi (2000)]{kobayashi}
{{Kobayashi}, S.},
2000, {\apj},
545,
807

\bibitem[ Kompaneets(1960)]{komp60}  Kompaneets A.S., 1960, Sov. Phys. Doklady, 130, 1001


\bibitem[{Kouveliotou} {\it et~al.}(1993)]{kmf+93}
{Kouveliotou}, C., {Meegan}, C.~A., {Fishman}, G.~J., {Bhat}, N.~P., {Briggs},
  M.~S. {\it et al.}s, 1993 \apjl, 413, 101 

\bibitem[Kumar \& Panaitescu(2000)]{Kumar00}
Kumar, P., \& Panaitescu, A. 2000, ApJ Lett., 541, 51 

\bibitem[Landau \& Lifshits(1982)]{LLVII}
{{Landau}, L.~D. and {Lifshits}, E.~M.},
1982, "{The electrodynamics of continuous media (2nd revised and enlarged edition)}", Pergamon Press



\bibitem[Larrabee \etal(2002)]{llm02}
{{Larrabee}, D.~A., {Lovelace}, R.~V.~E., {Romanova}, M.~M.
        }, 2003, {\apj},  586, 72 

\bibitem[Lazzati \& Begelman(2005)]{Lazzati05}
{{Lazzati}, D. and {Begelman}, M.}, 2005, astro-ph/0511658

\bibitem[Leblanc \& Wilson(1971)]{lw71}
{{Leblanc}, J.~M., {Wilson}, J.~R.}, 1970,  {\apj}, 161,
 541 


\bibitem[Liang \etal(2003)]{Liang} {{Liang}, E., {Nishimura}, K.},
2003,  arXiv:astro-ph/0308301

\bibitem[Lipunov \etal(2001)]{Lipunov}
{{Lipunov}, V.~M. and {Postnov}, K.~A. and {Prokhorov}, M.~E.
	}, 2001, {Astronomy Reports}, 45, 236

\bibitem[Lyutikov \& Usov(2000)]{lu00}
{{Lyutikov}, M. \& {Usov}, V.~V.}, 2000, 
{\apjl}, 543, 129


\bibitem[Lyutikov(2002a)]{lyu02}Lyutikov, M., 2002, Phys. Fluids, 14, 963 

\bibitem[Lyutikov(2003)]{l03}
{Lyutikov}, M., 2003, MNRAS, 346, 540


\bibitem[Lyutikov \& Blandford(2003)]{lb03}
Lyutikov M., Blandford R., 2003, astro-ph/0312347



\bibitem[Lyutikov \& Uzdensky(2003)]{lu03}  Lyutikov, M., Uzdensky D., 2003,
ApJ, 589, 893

\bibitem[Lyutikov \etal(2003)]{lyu03c} Lyutikov, M., Pariev, V., Blandford, R., 2003,
 ApJ,  597, 998

\bibitem[Lyutikov(2004)]{lyut04}
 Lyutikov, M., {35th COSPAR Scientific Assembly},
 p237, astro-ph/0409489

\bibitem[MacFadyen \etal(2001)]{mcfad01}
{{MacFadyen}, A.~I. and {Woosley}, S.~E. and {Heger}, A.},
2001, {\apj},
550, 410

\bibitem[McKinney \& Gammie(2004)]{Gammie} {{McKinney}, J.~C. and {Gammie}, C.~F.},
2004, 
ApJ, 611, {977-995}

\bibitem[McMahon \etal(2005)]{mcmahon}
{{McMahon}, E. and {Kumar}, P. and {Piran}, T.},
2005, astro-ph/0508087

\bibitem[Medvedev \& Loeb(1999)]{med99}
{{Medvedev}, M.~V. and {Loeb}, A.},
1999, {\apj}, 526, 697

\bibitem[Nakar \& Piran(2002)]{Nakar02}
{{Nakar}, E. and {Piran}, T.},
2002, MNRAS, 330, 920

\bibitem[Nakar \& Piran(2004)]{nakar}
{{Nakar}, E. and {Piran}, T.},
2004,
{\mnras},
 353,
 647

\bibitem[Nousek \etal(2005)]{Nousek}
{Nousek}, J.~A. \etal, 2005, astro-ph/0508332 

\bibitem[Piran (2004)]{Piran04} {{Piran}, T.}, 2004, 
The Physics of Gamma-Ray Bursts, astro-ph/0405503


\bibitem[Prochaska \etal(2005)]{Prochaska05}
Prochaska, J.X.,  \etal 2005, submitted to ApJL, astro-ph/0510022

\bibitem[Proga \etal(2003)]{prog03}
 {{Proga}, D., {MacFadyen}, A.~I., {Armitage}, P.~J., 
	{Begelman}, M.~C.}, 2003, astro-ph/0310002

\bibitem[Rees \& Gunn(1974)]{rg74} {{Rees}, M.~J., {Gunn}, J.~E.},
1974, {\mnras},
167,
1

 \bibitem[Retter \etal(2005)]{Rett05} Retter, A., Barbier. L., Barthelmy, S., et al.  2005, GCN Circ. 3788 


\bibitem[Rossi \etal(2002)]{ros02}
 {{Rossi}, E., {Lazzati}, D., {Rees}, M.~J.}, 2002, {\mnras}, 332, 945 

\bibitem[Rosswog  \etal(2003)]{ross03}
{{Rosswog}, S. and {Ramirez-Ruiz}, E. and {Davies}, M.~B.},
2003,
{\mnras},
345,
1077

\bibitem[Ruffert  \& {Janka}(2001)]{Ruffert}
{{Ruffert}, M. and {Janka}, H.-T.},
2001, {\aap},
 380,
 544

\bibitem[Rutledge \& Fox(2003)]{rutl03} Rutledge, R., Fox, D., 2003, astro-ph/0310385 

\bibitem[Sari \& Piran(1999)]{sari99}
{{Sari}, R. and {Piran}, T.},
1999,
 {\apj},
 520,
 641


\bibitem[Shapiro(1979)]{sha79} Shapiro, P., 1979, ApJ, 233, 831


\bibitem[Smolsky \& Usov(1996)]{su96} Smolsky~M.V., Usov~V.V. 1996, ApJ, 461, 858



\bibitem[{Stanek} {\it et~al.}(2003)]{smg+03}
{Stanek}, K.~Z., {Matheson}, T., {Garnavich}, P.~M., {Martini}, P., {Berlind},
  P., 2003, \apj,
       591, 17 

\bibitem[Tagliaferri\etal(2005)]{Tagliaferri}
{Tagliaferri}, G. \etal, 2005,
{\nat},
 436,
 985

\bibitem[Thompson \& Blaes(1998)]{tb98}
Thompson, C. \& Blaes, O. 1998, \prd, 57, 3219


\bibitem[Thompson \& Madau(2000)]{tm00}
{{Thompson}, C., {Madau}, P.}, 2000, {\apj}, 538, 105


\bibitem[Usov(1992)]{uso92}Usov, V. V., 1992, Nature, 357, 472

\bibitem[van Putten (2005)]{vanputten}
{{van Putten}, M.~H.~P.~M.},
2005,
astro-ph/0510348

\bibitem[Villasenor \etal(2005)]{Villasenor}
Villasenor, J.S., \etal, 2005, 
{\nat},
437, 
855 

\bibitem[Willis \etal(2005)]{Willis}
{{Willis}, D.~R. and {Barlow}, E.~J. and {Bird}, A.~J. and {Clark}, D.~J. and 
	{Dean}, A.~J. and {McConnell}, M.~L. and {Moran}, L. and {Shaw}, S.~E. and 
		{Sguera}, V.}, 2005,  {\aap}, 439, 
		245

\bibitem[Wheeler \etal(2005)]{Wheeler}
{{Wheeler}, J.~C. and {Akiyama}, S. and {Williams}, P.~T.}, 
2005, {\apss}, 298, 3

\bibitem[Woosley \etal(2003)]{Woosl02} {{Woosley}, S.~E., {Zhang}, W., {Heger}, A.},
2003,
in ''From Twilight to Highlight: The Physics of Supernovae'', 
p. 87


\bibitem[Woosley \&  Heger(2005)]{woosleyheger}
{{Woosley}, S. and {Heger}, A.},
2005, 
submitted to ApJ, astro-ph/0508175


\bibitem[Zenitani and Hoshino(2004)]{Hosh}
Zenitani, S. and  Hoshino, M., 2004,  astro-ph/0411373

\bibitem[Zhang \etal(2005)]{Zhang}
{Zhang}, B.  \etal, 2005, astro-ph/0508321


\end{thebibliography}
\end{document}